\documentclass[twocolumn,showpacs]{revtex4}
\usepackage{graphicx}
\usepackage{color}
\graphicspath{{pict/}{}}

\newcounter{Fig}

\newcommand{\be}{\begin{equation}}
\newcommand{\ee}{\end{equation}}

\begin{document}
\title{Diffusion in infinite and semi-infinite lattices with long-range coupling}
\author{Alejandro J. Mart\'inez and Mario I. Molina}
\address{Departamento de F\'isica, Facultad de Ciencias,
Universidad de Chile, Santiago, Chile and\\ Center for Optics and
Photonics (CEFOP), Casilla 4016,
Concepci\'on, Chile.}
\begin{abstract}
We prove that for a one-dimensional infinite lattice, with long-range coupling among sites, the diffusion of an initial delta-like pulse in the bulk, is ballistic at all times. We obtain a closed-form expression for the mean square displacement (MSD) as a function of time, and show some cases including finite range coupling, exponentially decreasing coupling and power-law decreasing coupling. For the case of an initial excitation at the edge of the lattice, we find an approximate expression for the MSD that predicts ballistic behavior at long times, in agreement with numerical results.
\end{abstract}

\maketitle



The physics of discrete systems have been a topic of interest for many years, because
it can give rise to completely different phenomenology in
comparison with that present in homogeneous continuous systems. In particular, discrete periodic systems are found in many different contexts including condensed matter physics, optics, Bose-Einstein condensates and magnetic metamaterials among others. Under the appropriate approximation, they can all be  described by some variant of the discrete Schr\"{o}dinger (DS) equation~\cite{rep}. In that way, many of these systems display the phenomenology common to periodic systems such as the presence of a band structure, discrete diffraction, Bloch oscillations, dynamic localization, Zener tunneling, to name a few.

Usually, the DS equation is used in the weak-coupling limit, where the interaction among sites includes nearest-neighbors only. This is a good approximation in cases where the coupling among sites decays very quickly with distance, like the exponentially-decreasing coupling found in optical waveguide arrays. However, there are cases where it is advisable to go beyond this approximation. An example of that is a split-ring resonator (SRR) array, where the interaction among the basic units is dipolar in nature and therefore, the coupling decreases as the inverse cubic power of the mutual distance.

When coupling beyond nearest-neighbors are considered, the number of possible routes of energy exchange increases. The dynamical evolution of excited pulses in finite 1D and 2D  lattices with anisotropic couplings and up to second nearest-neighbor couplings, has been explored in \cite{sza1} by means of the Green function formalism. Experimental observation of the influence of second order coupling in linear and nonlinear optical zig-zag waveguide arrays has been recently carried out~\cite{zz}. In a different 
context, a recent work~\cite{transition}, shows that long-range coupling in
low dimensional system can induce a phase transition from delocalized to
localized modes. Another interesting scenario that can be modeled as a discrete
system with long-range couplings is that of complex networks~\cite{networks}, 
where the distances between nodes are not necessarily physical. 

In this Letter we carry out an analytical and numerical study on the 
diffusion of an initially localized pulse propagating in a one-dimensional discrete periodic lattice, in the presence of arbitrary long-range couplings. We focus on two cases of interest: A delta-like pulse in the bulk, and a delta-like pulse at the edge of the lattice. The analytical work centers on the evaluation of the mean square displacement of the excitation, which is obtained in exact form for the bulk excitation, and in an approximate form for the edge excitation.


Let us consider the (dimensionless) discrete Schr\"{o}dinger(DS) equation, describing the evolution of an excitation along a one-dimensional periodic lattice:
\begin{equation}
i\frac{du_n}{dz}+\sum_{m\neq n}V_{n,m}u_m=0
\label{eq1}
\end{equation}
where $u_n$ is the complex amplitude of the excitation at the $n$th
site, $z$ is the evolution coordinate ('time' in the tight-binding model for electrons, or 'longitudinal distance' for coupled waveguide arrays in optics ). Matrix 
element $V_{n,m}$ denotes the coupling between the $n$th and $m$th sites. This matrix
is periodic in space and obeys $V_{n,m}=V_{m,n}=V_{|n-m|}$. Eq.
(\ref{eq1}) conserves the norm $P =
\sum_{-\infty}^{\infty}|u_n(z)|^2$ which can then be set, without loss of generality, as unity, $P=1$. 
The dispersion relation of the linear waves is obtained by inserting a solution
of the form $u_n=Ae^{i(kn+\lambda z)}$ in Eq. (1), obtaining
\begin{equation}
\lambda=\sum_{m\neq n}V_{n,m}e^{ik(m-n)}=2\sum_{m=1}^{\infty}V_m\cos(mk).
\label{disp}
\end{equation}
The convergence of this series for all $k$ values constrains $V_{n}$ to decrease faster than $1/n$. From Eq.(\ref{disp}) we immediately obtain some basic
properties of $\lambda(k)$:\ $\lambda(k)=\lambda(-k)$, $\lambda(k)=\lambda(k+2\pi q),\;q\in Z$, and $\partial_k\lambda(k)|_{k=0} = \partial_k\lambda(k)|_{k=\pm\pi}=0$.

The dynamical evolution of an initially localized  pulse in a lattice, $u_n(0)=A_0\delta_{n,n_{0}}$, can be monitored through the
mean square displacement(MSD) of the excitation:
\be
\left<n^2\right>\equiv\sum_{M}^{\infty}n^2|u_n(z)|^2/
\sum_{M}^{\infty}|u_n(z)|^2,\label{msq_bulk}
\ee
\begin{figure}[t]
\centering
\includegraphics[width=8.5cm]{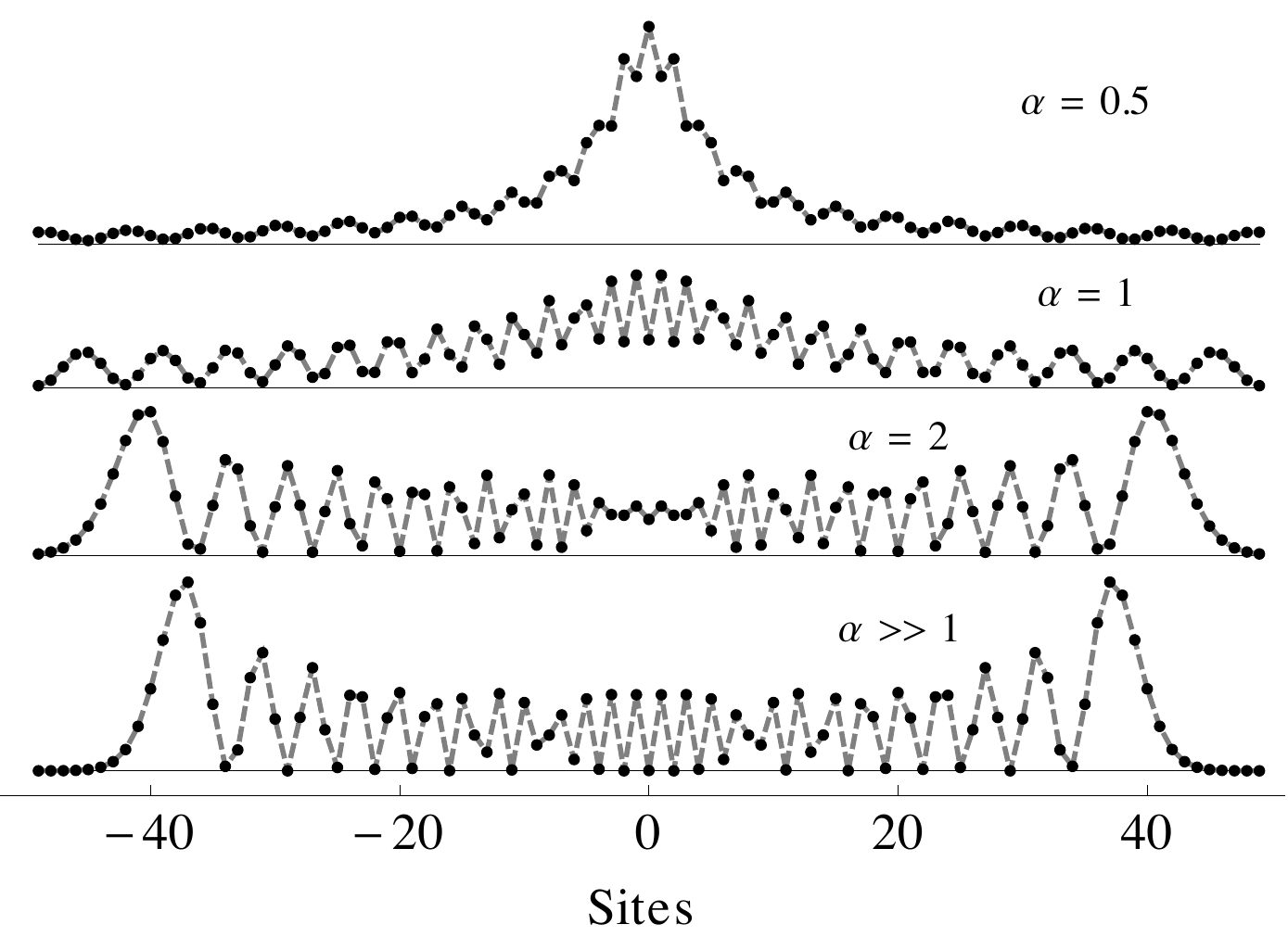}
\caption{Discrete diffraction pattern for a bulk spread of a
delta-like initial condition at $Vz=20$ for different values
of dispersion parameter $\alpha$ in a lattice with coupling exponentially
decreasing.}
\label{fig1}
\end{figure}
where $|n_{0}|\gg 0$ and $M=-\infty$ for an infinite lattice, or $n_{0}=0$ and $M=0$ for a 
semi-infinite lattice.


{\em Diffusion in the bulk}. In this case $u_n(0)=A_0\delta_{n,n_{0}}$, where $n_{0}$ is 
far away from the boundaries of the lattice. By combining 
Fourier-and-Laplace transforms to Eq. (1), followed by the 
corresponding transformation back to space $n$ and $z$
coordinates, one obtains a formal expression for $u_{n}(z)$
\begin{equation}
u_n(z)=\frac{A_0}{2\pi}\int_{-\pi}^{\pi}e^{i(kn-\lambda(k)z)}dk
\label{generalsol}
\end{equation}
where $\lambda(k)$ is the dispersion relation. Next we proceed
to insert Eq.(\ref{generalsol}) into Eq.(\ref{msq_bulk}), and after
using the general properties of $\lambda_{k}$, one obtains after some algebra:
\begin{equation}
\left<n^2\right> = \left[\frac{1}{2\pi}\int_{-\pi}^{\pi}\left(\frac{d\lambda(k)}{dk}
\right)^2dk\right]z^2.
\label{msd}
\end{equation}
Equation (\ref{msd}) implies that the propagation of the excitation is 
{\it ballistic at all times}, with a 'speed' that depends upon the 
smoothness of the dispersion relation.  This result is also valid in any dimension $d$,
where it can be easily proven that
\begin{equation}
\left<{\bf n}^2\right> = \left[\frac{1}{v}\int_{FBZ}\left(\nabla_k\lambda(k)
\right)^2d^dk\right]z^2
\label{msd2}
\end{equation}
where the integral is taken over the first Brillouin zone (FBZ), with
volume $v$. An equivalent expression to Eq. (\ref{msd2}) in ``real space'' is
obtained by replacing (\ref{disp}) in (\ref{msd2})
\begin{equation}
\left<{\bf n}^2\right>=\left[\sum_{\bf m}{\bf m}^2|V_{\bf 0,m}|^2\right]z^2
\label{msd3}
\end{equation}
where ${\bf m}$ is the relative position if a lattice node from an
arbitrary site ${\bf n}$ taken here as ${\bf 0}$ without loss of
generality because of periodicity, and $V_{\bf 0,m}$ is the coupling
between sites ${\bf 0}$ and ${\bf m}$, where $V_{\bf n,m}$ is periodic
in space and obeys: $V_{\bf n,m}=V_{\bf -n,-m}$.

Let us compute now in detail the MSD for several cases of interest.

(a)\ {\em Second-order coupling}. 
In this case, $V_1=V$, $V_2=\beta V$ and $V_{i>2}=0$, and the dispersion relation is
$\lambda = 2V(\cos(k)$ $+\beta\cos(2k))$. This implies, according to Eq.(\ref{msd})
\begin{equation}
\left<n^2\right>=2(1+4\beta^2)(Vz)^2.
\end{equation}
Thus, the inclusion of a second-order coupling (only) always increases the
speed of the transversal diffusion.

(b)\ {\em Exponentially-decreasing coupling}.\ In this case, $V_{n,m}=Ve^{-\alpha(|n-m|-1)}$, where $V$ is the coupling
between nearest-neighbors sites and
$\alpha$ is the long-range parameter. 
\begin{figure}[t]
\centering
\includegraphics[width=8.cm]{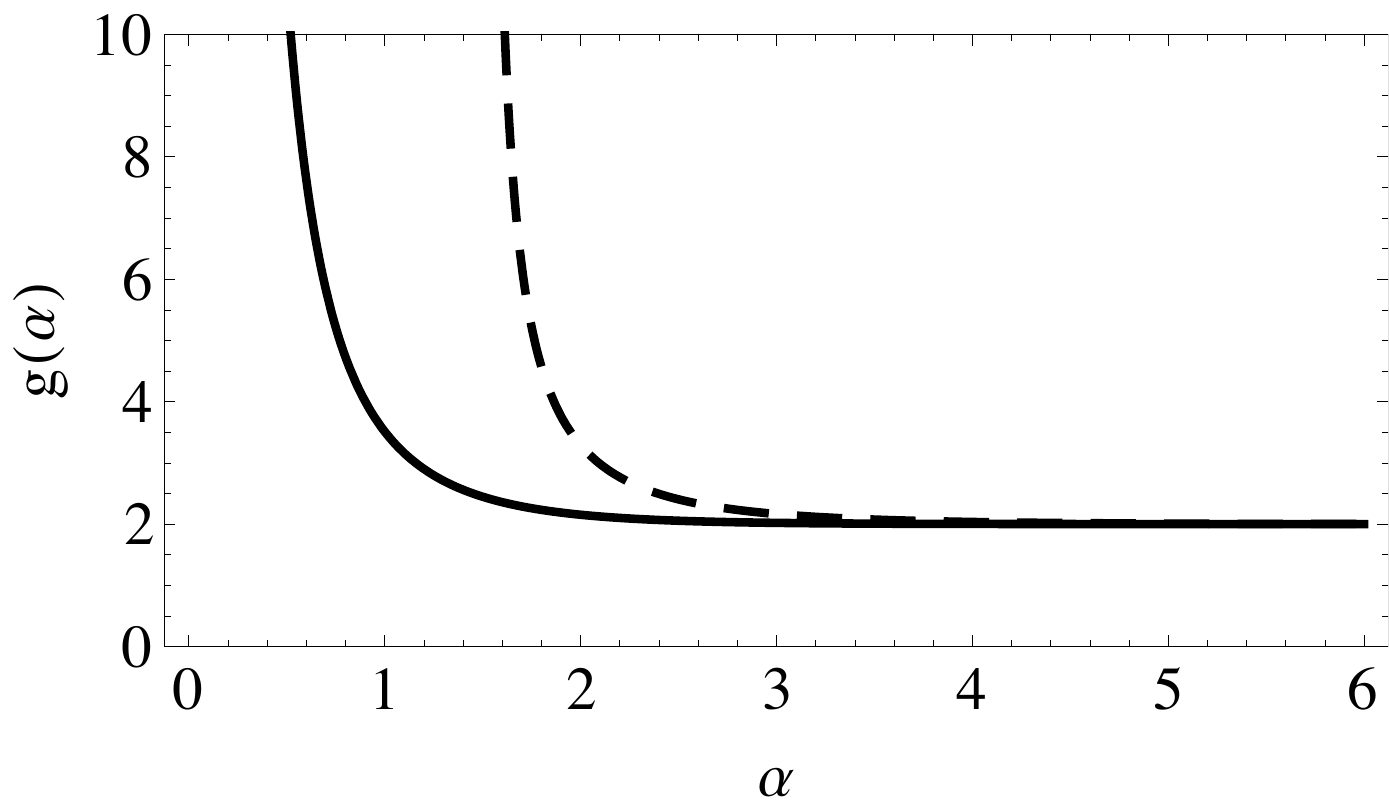}
\caption{Speed of ballistic propagation of initially
localized excitation, as a function of dispersion parameter. The solid
(dashed) curve corresponds to exponential (power-law) coupling case.
For power-law coupling $g(\alpha)=2\zeta(2(\alpha-1))$.}
\label{fig2}
\end{figure}
The linear dispersion relation is
\begin{equation}
\lambda=V\left(\frac{e^{\alpha}\cos(k)-1}{\cosh(\alpha)-\cos(k)}\right)
\label{displr}
\end{equation}
Figure~\ref{fig1} shows some snapshots of the 
spatial transversal profiles at a
given longitudinal propagation distance $z$, for different values of the dispersion
parameter. We note that, at small $\alpha$ values, a fraction of the
initial excitation seems to `linger' at the initial 
position\cite{sza1,complex}, in
a short of quasi-selftrapping, while the ``untrapped'' portion
propagates away from the initial site faster than in the nearest-neighbor
case. These observations can be put on a more rigorous basis
by computing the mean square displacement directly from Eq. (\ref{msd}),
using Eq. (\ref{displr}). We obtain:
$\left<n^2\right>=g(\alpha)(Vz)^2$ where,
\begin{equation}
g(\alpha) = \frac{1}{2}\coth(\alpha)(\coth(\alpha)+1)^2.
\label{g}
\end{equation}
This implies a ballistic speed $g(\alpha)\geq 2$, for all
finite $\alpha$. However, it only becomes noticeable larger than two
for $\alpha\leq 2$ (Fig. ~\ref{fig2}). 

(c)\ {\em Power-law coupling}.\ Another popular case of long-range interaction is the power-law coupling $V_{n,m} =
V/|n-m|^\alpha$.
In this case the dispersion relation is given in terms of a 
polylogarithm function
$\lambda(k,\alpha)=V\left(Li_{\alpha}(e^{ik})\right.$
$\left.+Li_{\alpha}(e^{-ik})\right)=
2V\sum_{m=1}^{\infty}\cos(mk)/m^\alpha$.
For $\alpha=1$ we can explicitly written as
\begin{equation}
\lambda(k,1)=-V\log\left|2-2\cos(k)\right|
\end{equation}
with the logarithm diverges for $k=0$. On the other hand, 
for $\alpha>1$ the polylogarithm remains bounded at
$|k|\leq\pi$. Moreover, in this case we 
can use Eq. (\ref{msd3}) to easily calculate
the mean square displacement, obtaining
\begin{equation}
\left<n^2\right> =2\zeta(2(\alpha-1))(Vz)^2
\end{equation}
where $\zeta$ is the Riemann zeta function. Figure \ref{fig2} shows
the ``speed'' of diffusion as a function of $\alpha$ and compares it with the
``speed'' obtained for lattices with exponentially-decreasing coupling.
The discrete diffraction pattern (Fig.1) observed in case (b), is also observed in this case.

In all three cases examined, the speed of diffusion is greater than in the case with coupling to nearest-neighbors only. For the exponential and power-law cases, we note that as the range of the interaction is increased, the pulse experiences a sort of quasi-localization at the initial site. At the same time the speed $g(\alpha)$ increases above $2$ and tends to diverge at $\alpha=0$ (Fig.2). We can understand this quasi-localization phenomenon by analyzing the dispersion
relations more closely. In cases (b) and (c), as the range increases, the
dispersion gets flatter and flatter in the vicinity of $=\pm \pi$, signaling the 
emergence of linear modes with very small group velocities. As $\alpha$ is increased further, this vicinity grows and most of the linear modes acquire negligible velocity, with the exception of a small vicinity of $k=0$, where the concavity is very high, giving rise to long wavelength modes with high velocity. Since our initial delta-like condition is a superposition of all these linear waves, the pseudo-localization phenomena can be understood as due to those modes with zero velocity, while those few and fast long wavelength modes, give rise to the wings that escape at high speed.

In the formal limit $\alpha=0$ and for a finite number of sites $N$, all sites are connected to one another, and 
the high degree of degeneracy forces the localization of 
the wave function through the execution of incomplete oscillations between the initial
excited site and all the other ones. In this case the effective dynamics can be mapped to the dynamics of an asymmetric dimer\cite{tetra}. This system has been solved in closed form and shows partial linear localization at the initial site for finite $N$, which becomes complete in the limit $N\rightarrow \infty$\cite{complex}.
  
{\em Diffusion at the boundary}. In this case $u_n(0)=A_0\delta_{n,0}$, i.e., the
initial excitation is at the very edge of the lattice. This system can be viewed as an infinite lattice with boundary conditions: $u_n(z)=0$ for $n<0$, for all $z$. The lack of translational symmetry prevent us from writing a general expression for $u(z)$, similar to  Eq.(\ref{msd}). However, the case of nearest-neighbors coupling only, can be solved in closed form using the method of images\cite{images}.
The boundary conditions are obeyed if we placed an opposite delta-like source at $n=-2$. Then, the propagation for $n\geq0$ is then the superposition of the evolution of both sources
\begin{figure}[t]
\centering
\includegraphics[width=8.5cm]{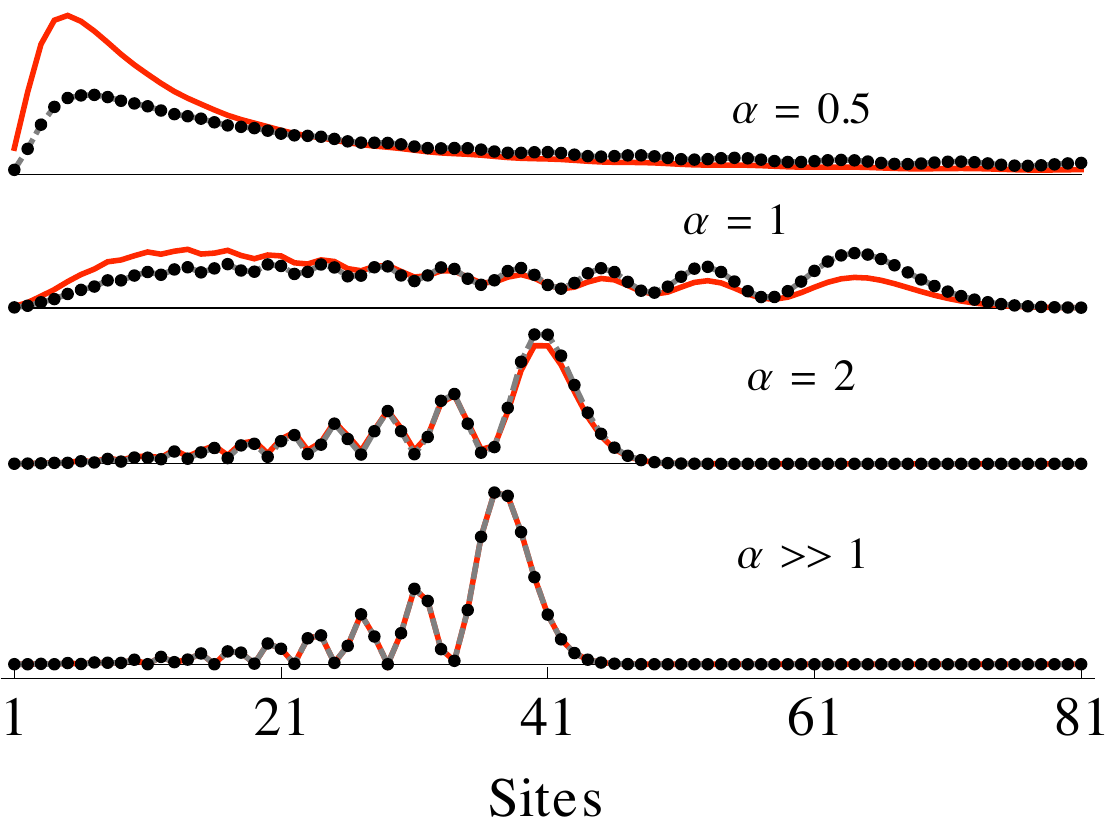}
\caption{Discrete diffraction pattern for a surface spread of a
delta-like initial condition at $Vz=20$ for different values
of dispersion parameter $\alpha$ in a lattice with exponentially-decreasing 
coupling. The solid line corresponds to approximation (\ref{solsup}).}
\label{fig3}
\end{figure}
\begin{eqnarray}
u_n^s(z)&=&u_{n}(z)-u_{n+2}(z)\nonumber\\
&=&\frac{A_0}{2\pi}\int_{-\pi}^{\pi}\left(1-e^{-2ik}\right)
e^{i(kn-\lambda(k)z)}dk,
\label{solsup}
\end{eqnarray}
where index $s$ denotes a semi-infinite
array. However, When long-range couplings are considered, it is no longer possible to satisfy the boundary conditions by the images method. The reason has to do with the fact that, even though we have $u_{-1}(z) = 0$ for all $z$, the fields of each semi-infinite arrays are still coupled by the range of the interaction.

However, by comparing the predictions of Eq.(\ref{solsup}) with numerical results for the dynamical evolution, we find that Eq.(\ref{solsup}) still provides a good approximation to the general case, by just substituting the corresponding dispersion relation for a given long-range coupling. After using general properties of the dispersion relation, one obtains an asymptotic expression
\begin{equation}
\left<n^2\right>^s \approx \left[\frac{1}{\pi}
\int_{-\pi}^{\pi}\sin^2(k)\left(\frac{d\lambda(k)}{dk}
\right)^2dk\right]z^2,
\label{msdsur}
\end{equation}
valid for $z\gg 1$.
\begin{figure}[t]
\centering
\includegraphics[width=7.cm]{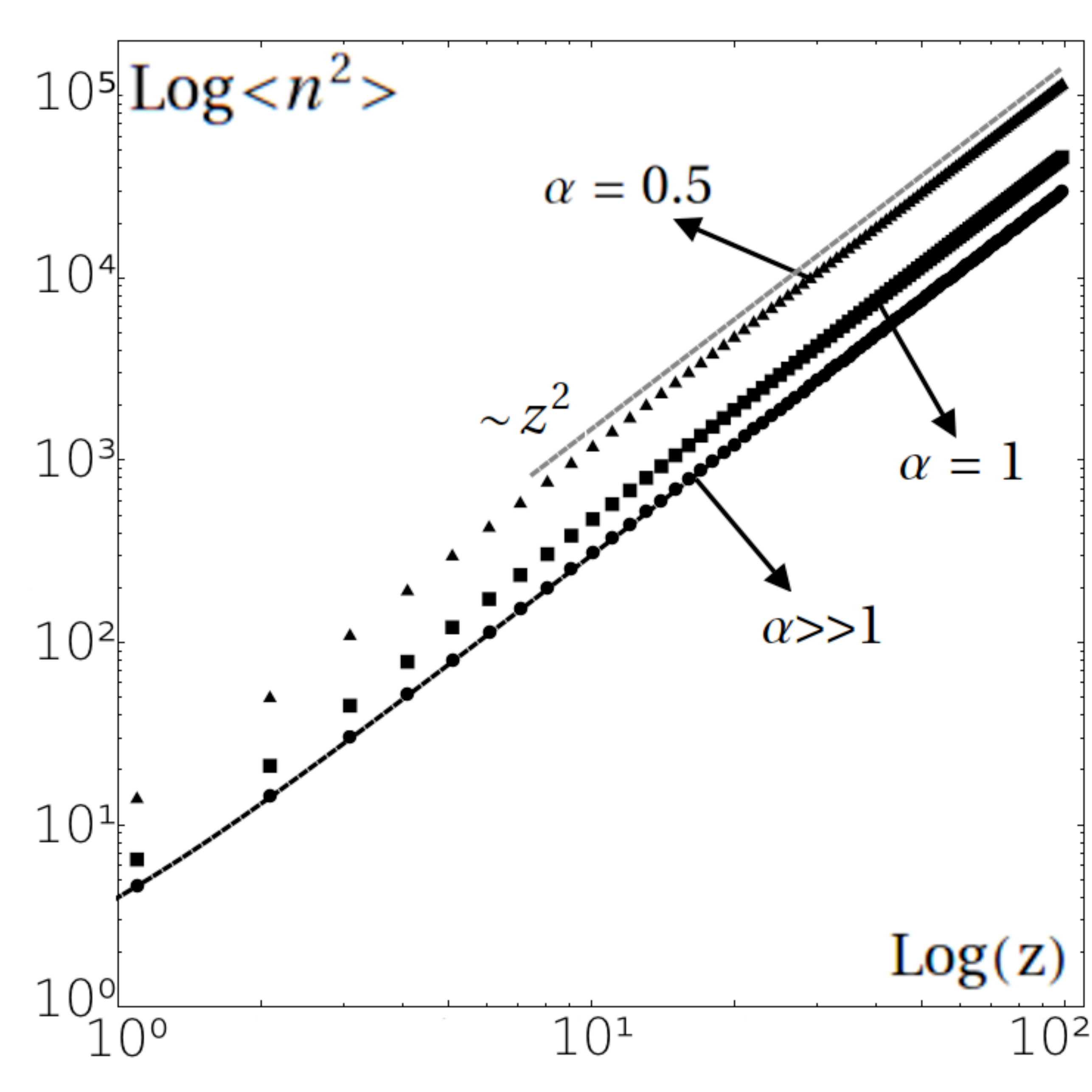}
\caption{MSD in logarithmic scale for exponentially-decreasing coupling. 
Black line shows the approximate solution obtained using the
method of images, Eq.(\ref{solsup}), while the dashed line shows the slope associated to $z^2$ (to guide the eyes only).}
\label{fig4}
\end{figure}

\begin{figure}[t]
\centering
\includegraphics[width=8cm]{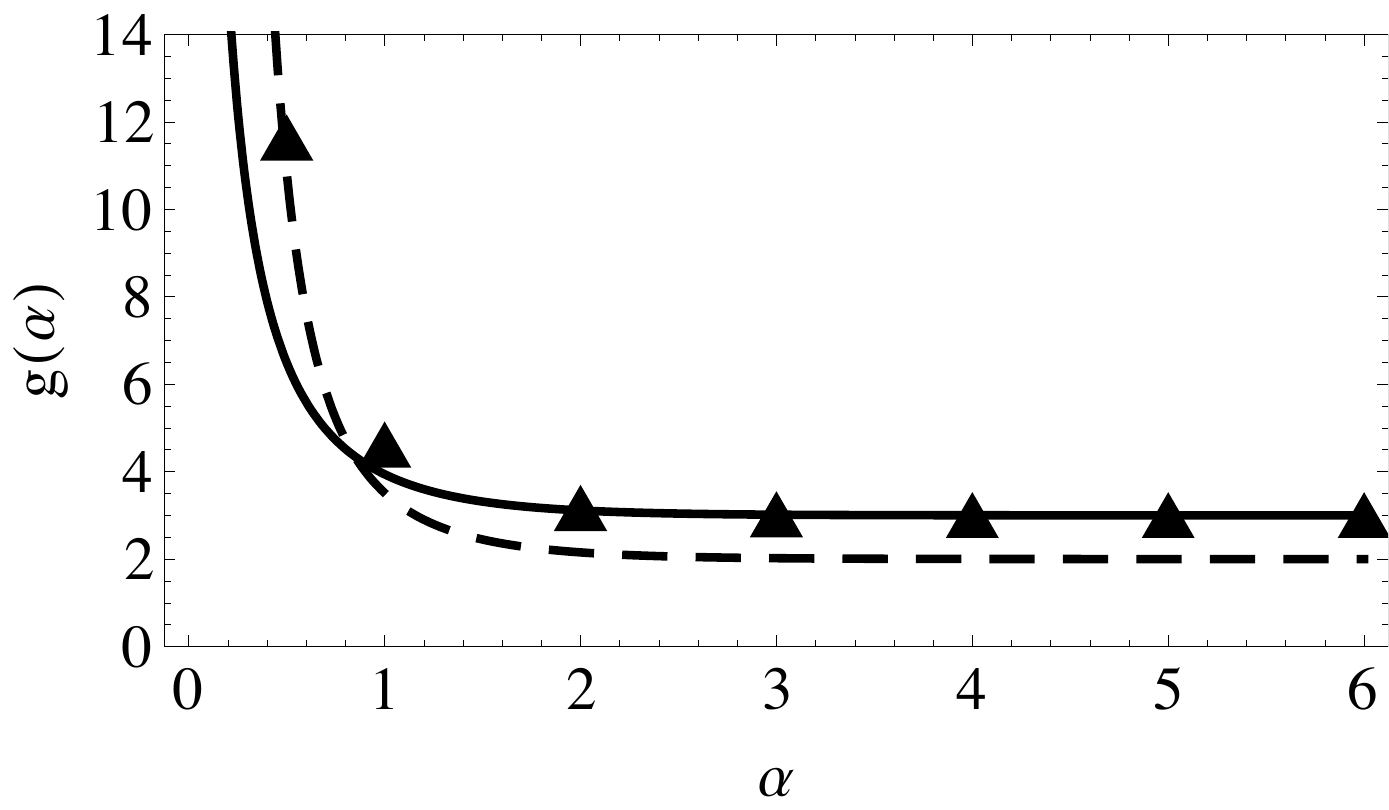}
\caption{Square of speed of ballistic propagation of initially
localized excitation, as a function of dispersion parameter $\alpha$ for 
the exponentially-decreasing coupling. Triangles correspond to numerical values, continuous line to approximation (\ref{approx})
and dashed line to the exact bulk solution, Eq. (\ref{g}).}
\label{fig5}
\end{figure}

As expected, the asymptotic behavior is ballistic in all cases, after some transient time, related to the repulsive effect of the boundary. Equation (\ref{msdsur}) can also be expressed in real space, by replacing Eq.(\ref{msd3}) in Eq. (\ref{msdsur}), obtaining
\begin{equation}
\left<n^2\right>^s \approx
\left[3V_1+2\sum_{m=2}^{\infty}m^2V_m^2
-2\sum_{m=1}^{\infty}m(m+2)V_mV_{m+2}\right]z^2.
\end{equation}
This results are confirmed by direct computations of cases (a), (b) and (c), using `exact' Eq.({msdsur}) with an edge excitation as an initial condition, and comparing it with the asymptotic formulas. Fig.\ref{fig4} shows this comparison of the MSD for the exponentially-decreasing coupling case. In this case, the asymptotic behavior is
\begin{equation}
\left<n^2\right>^s\approx 3\coth(\alpha)(Vz)^2.
\label{approx}
\end{equation}
In Fig. (\ref{fig5}) we show a comparison between this approximation 
and the numerical results. 
We see that for $\alpha\gtrsim 1$ there is a good correspondence between the numerical and analytical approach, while 
for $\alpha\leq 1$, the exact solution for the bulk, Eq. (\ref{g}), 
constitutes a better  fit than the surface approximation, Eq.(\ref{approx}). This is understandable because in the long-range coupling regimen, the  boundary loses its meaning as such and the edge site behaves effectively like a bulk site.


In conclusion, we have examined the propagation of an excitation initially placed in the bulk and at the boundary of a one-dimensional lattice with long-range coupling. For the bulk case, we find a closed-form expression showing that the propagation is ballistics at all times, while for the edge excitation, we obtain an asymptotic approximation based on the method of images, that predicts ballistic propagation at long times, in agreement with numerical results.

This work was supported in part by FONDECYT Grants 1080374 and
1070897, and Programa de Financiamiento Basal de CONICYT (Grant
FB0824/2008).\\

\end{document}